\documentclass[aps,prb,twocolumn]{revtex4}
\usepackage{color,graphics}
\newcommand{\figwidth}{3.375in}
\begin{document}
\preprint{ISSP}
\title[Short Title]
{Dynamical Critical Phenomena in
three-dimensional Heisenberg Spin Glasses }
\author{Mitsuhiro Matsumoto}
\author{Koji Hukushima}
\email{hukusima@issp.u-tokyo.ac.jp}
\author{Hajime Takayama}
\affiliation{
Institute for Solid State Physics, University~of~Tokyo, 5-1-5 Kashiwa-no-ha,
Kashiwa, Chiba 277-8581, Japan}
\date{\today}
\begin{abstract}
Spin-glass (SG) and chiral-glass (CG) orderings in three dimensional
 ($3D$) 
 Heisenberg spin glass with and without magnetic anisotropy are studied
 by using large-scale off-equilibrium Monte Carlo simulations. 
A characteristic time of relaxation, which diverges at a transition
 temperature in the thermodynamic limit, is obtained as a function of
 the temperature and the system size. Based on the finite-size scaling analysis
 for the relaxation time, it is 
 found that in the isotropic Heisenberg spin glass, the  CG
 phase transition occurs at a finite temperature, while the SG
 transition occurs at a lower temperature, which is compatible with
 zero. 
Our results of the anisotropic case  support the chirality scenario for
 the phase transitions in the $3D$ Heisenberg spin glasses. 
\end{abstract}
\maketitle

\section{\label{sec:introduction}Introduction}
Chirality in frustrated magnets, which provides fascinating
magnetic orderings and/or novel critical phenomena, has attracted
interest in recent years\cite{Chirality}.
In the spin-glass (SG) study, the chirality-driven mechanism has 
been proposed by Kawamura\cite{Kawamura92} in order to explain the
origin of experimentally observed SG phase transitions.  
The chirality scenario assumes that a chiral-glass (CG) long-range order exists at
low temperatures with preserving the proper rotational symmetry in a
fully isotropic SG model.
By random magnetic anisotropy even of a small magnitude, which always
exists in real experimental situations, a SG phase is expected to emerge
as a result 
that the anisotropy mixes the two degrees of freedom, spin and
chirality. 
Therefore, in the chirality-driven mechanism the SG phase transition
experimentally observed in a class of compounds such 
as CuMn is essentially governed by the CG fixed point. 

Some numerical results on equilibrium properties support that 
the chirality-driven phase transitions occur 
in a three dimensional ($3D$) Edwards-Anderson (EA) Heisenberg SG
model\cite{Kawamura98,HukushimaKawamura00,KawamuraImagawa01} as well as in
a $3D$ XY SG model\cite{Li}.  
The CG transition temperature and the critical exponents are
estimated. They seem to be compatible with those observed
experimentally\cite{Campbell}.
However, some researchers have argued the possibility of a
finite-temperature SG transition\cite{Matsubara01}, or even simultaneous
SG and CG transition in the $3D$ Heisenberg SG
model\cite{Nakamura-Endoh01}. In a $3D$ XY SG model, large scale
domain-wall renormalization group calculation at zero temperature
suggests that a SG long-range order occurs at non-zero temperature below
the CG transition temperature\cite{Maucourt}. A similar study obtained
by different boundary conditions also support the existence of SG order
at finite temperatures\cite{Kosteritz-Akino}.  

The observed critical behavior associated with the CG
transition is found to be remarkably different from that observed in 
ordinary second-order phase transitions. For instance, the crossing of
the Binder parameter of systems with different sizes at the transition
temperature, which is usually observed in standard second-order
transitions, never occurs in this
model\cite{HukushimaKawamura00,KawamuraImagawa01}.
Furthermore, the previous study revealed that the CG phase of the $3D$
isotropic Heisenberg EA model had a peculiar feature which was
consistent with one-step replica symmetry breaking (RSB)
phase\cite{HukushimaKawamura00}. Some mean-field SG models with the
one-step RSB exhibit dynamical singularity  at a temperature well above
the static transition temperature\cite{Review}. 
Recent simulation on a mean-field SG model\cite{Brangian} shows that
dynamical finite-size scaling holds for the dynamical transition
temperature which is different from the static one .  
It is of particular interest to clarify whether the CG phase of the $3D$
Heisenberg EA model belongs to this class or not.  

Until now,  there have been a few works on  dynamical properties
of the Heisenberg SG models. 
Olive et al\cite{Olive86} and Yoshino and one of the authors
(HT)\cite{Yoshino93} 
investigated spin autocorrelation functions in {\it equilibrium} by
Monte Carlo simulations. They found that, with the help of
dynamical scaling analysis, the temperature dependence of the
characteristic relaxation time
is compatible with a zero-temperature SG transition. However, the
corresponding CG relaxation time has not been investigated yet. 

In the present paper, we study dynamical properties associated both with 
spin and chiral degrees of freedom above the 
transition temperature of the $3D$ Heisenberg EA model with and
without anisotropic interactions.
We developed an efficient method for extracting the equilibrium
relaxation time from non-equilibrium simulation for a given finite-size
system and analyzed the time scale by using the finite-size scaling. 
The results of our analyses on the isotropic Heisenberg SG model support
the first assumption of the chirality mechanism, namely the existence of
a finite temperature CG phase transition without a conventional SG
long-range order. 
In particular, 
a crossover from high-temperature SG dominant regime to the CG dominant
regime is clearly observed with decreasing temperature. 
In addition,  by introducing a finite random anisotropic interaction,
the SG relaxation time is found to be significantly 
enhanced at low temperatures and in systems with large sizes, 
while there is no such drastic change
in the CG dynamics. This is consistent with the chirality mechanism. 

Our work
is distinct from the previous works in the 
following senses;
(i) The CG dynamics as well as the
SG ones have been studied systematically both above and around a
transition temperature, 
(ii) The equilibrium relaxation is estimated from
non-equilibrium MC simulations.
These results may be complementary with the previous equilibrium
studies\cite{Kawamura92,HukushimaKawamura00}. 

The present paper is organized as follows. 
In section \ref{sec:model}, we explain the model and the method of our
Monte Carlo simulation. 
We introduce a dynamical ratio function  and discuss
its characteristic features in section \ref{sec:ratio} . 
Finite-size scaling of the characteristic time  is studied in section
\ref{sec:FSS} and section \ref{sec:summary}  is devoted to summary of
the paper. 

\section{Model and simulations}
\label{sec:model}
The model studied is a classical Heisenberg model on a
simple cubic 
lattice, defined by the Hamiltonian, 
\begin{equation}
 H(\vec{S})=-\sum_{\langle ij\rangle}J_{ij}\vec{S}_i\cdot\vec{S}_j - \sum_{\langle ij\rangle,\mu,\nu}D_{ij}^{\mu\nu}S_i^{\mu}S_j^{\nu}, 
\label{eqn:sec3-Model}
\end{equation}
where $\vec{S}_i=(S_i^x,S_i^y,S_i^z)$ is a three-component unit vector,
and the sum runs over all nearest-neighbor pairs with the total number
of sites $N=L\times L\times L$. The nearest neighbor couplings $J_{ij}$
are random bimodal 
variables which take $\pm J$ with equal probability. 
The anisotropic interactions $D_{ij}^{\mu\nu}$ are also independent
random variables obeying a box distribution with its range
$[-D:D]$. We assume that the anisotropic term is symmetric,
$D_{ij}^{\mu\nu}=D_{ij}^{\nu\mu}$.
A quite similar model has been studied by MC
simulation\cite{Matsubara91} with some values of strength $D$, where the
SG phase at low temperatures has been found around and above $D/J\sim
0.05$. We study the model system both with the isotropic case ($D/J=0$)
and an anisotropic case ($D/J=0.05$). 

We employ a standard heat-bath Monte Carlo method with two sub-lattice
flips\cite{Olive86}. One Monte Carlo step (MCS) is defined as $N$ spin trials. 
We perform MC procedure as follows; firstly spins are set as a random
configuration, namely they are instantaneously quenched from the high
temperature limit. The spin configuration is updated at a working
temperature $T$ during a waiting time $t_w$, which is from $10^0$ to $10^5$. 
Subsequently, additional MC steps are performed for measurements. 
The lattice sizes $L$ studied are $4$, $8$, $16$ and $32$. 
In our simulations, we perform eight independent runs for each identical 
sample to take an average over initial conditions and different thermal
noises.  A typical number of samples averaged over is 40--960 depending
on the system size.

\begin{figure}[]
\resizebox{\figwidth}{!}{\includegraphics{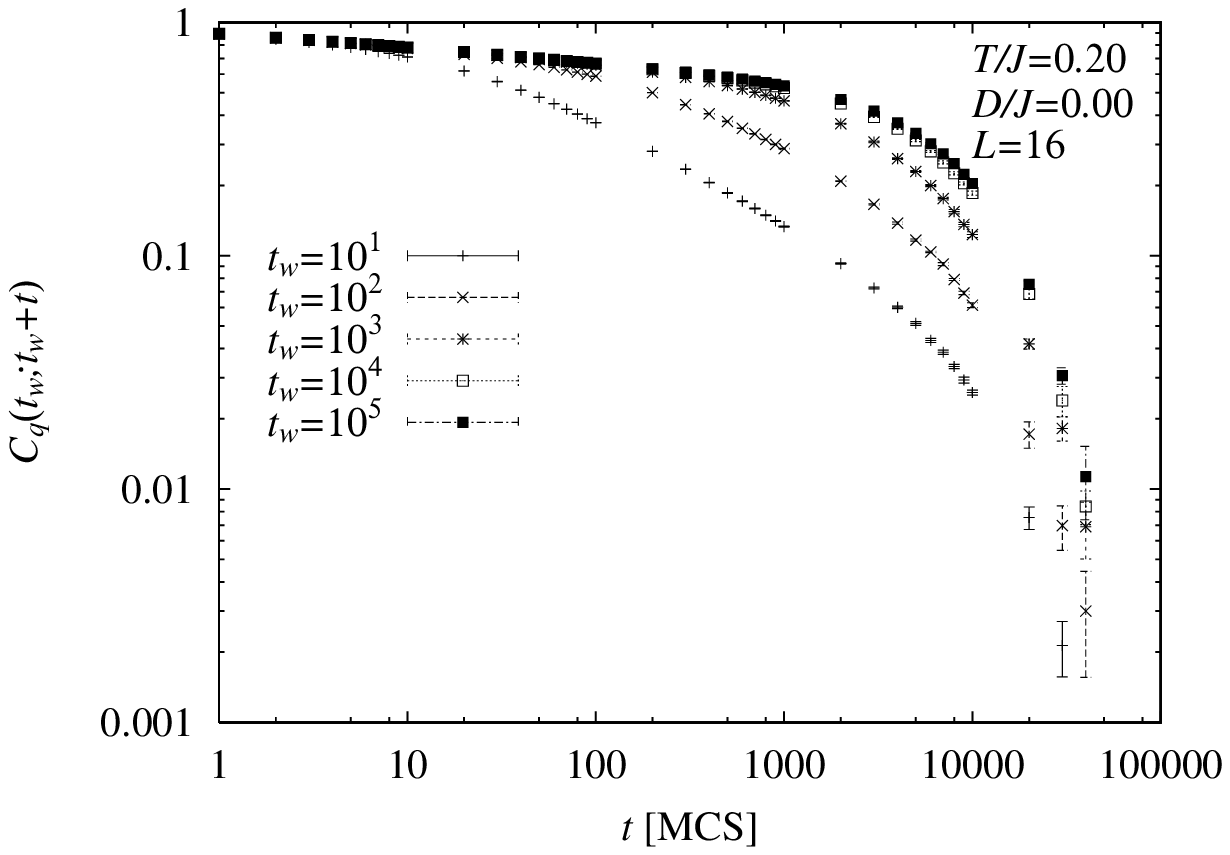}}
\resizebox{\figwidth}{!}{\includegraphics{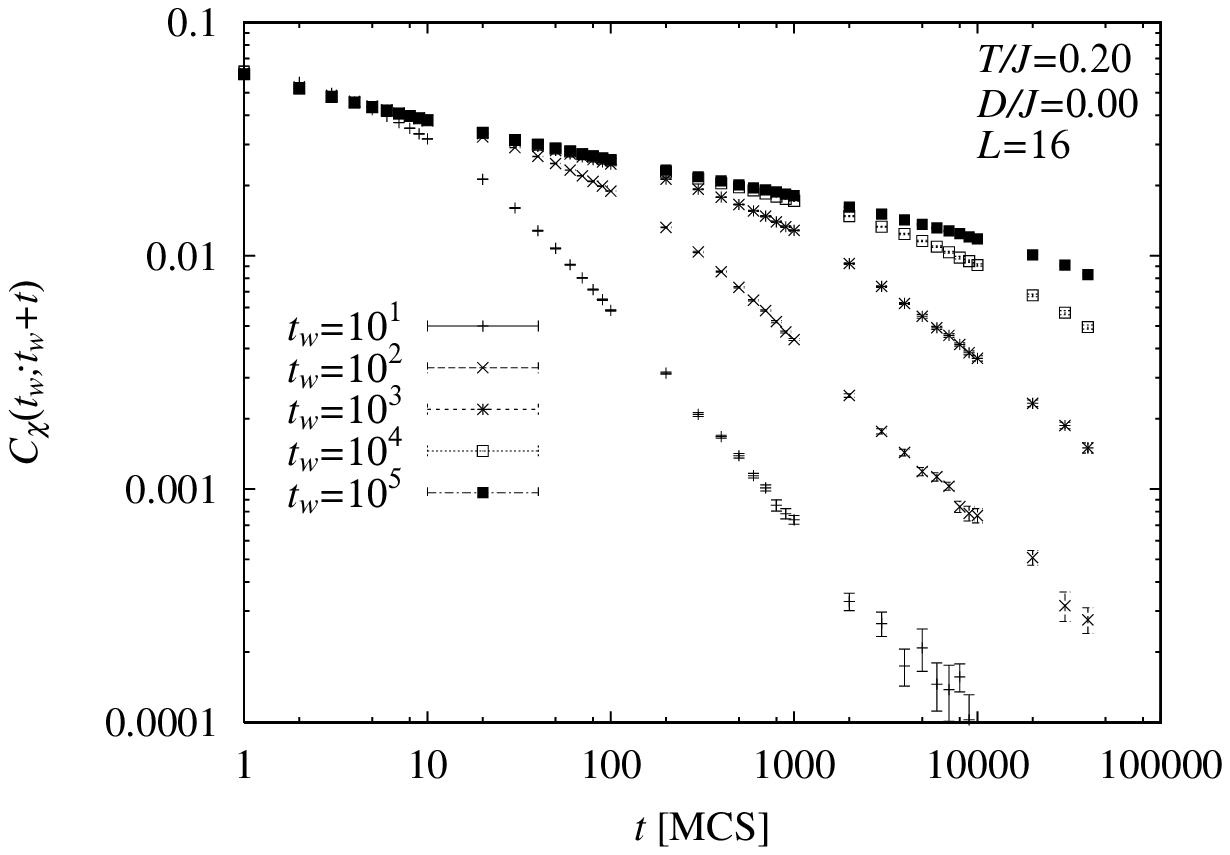}}
\caption{Time dependence of the spin-glass (upper) and chiral-glass
 (lower) autocorrelation functions with different waiting times in the
 $3D$ isotropic Heisenberg SG model at $T/J=0.20$.
The system size is fixed, $L=16$. 
} 
\label{fig3:rawq}

\end{figure}
\begin{figure}[]
\resizebox{\figwidth}{!}{\includegraphics{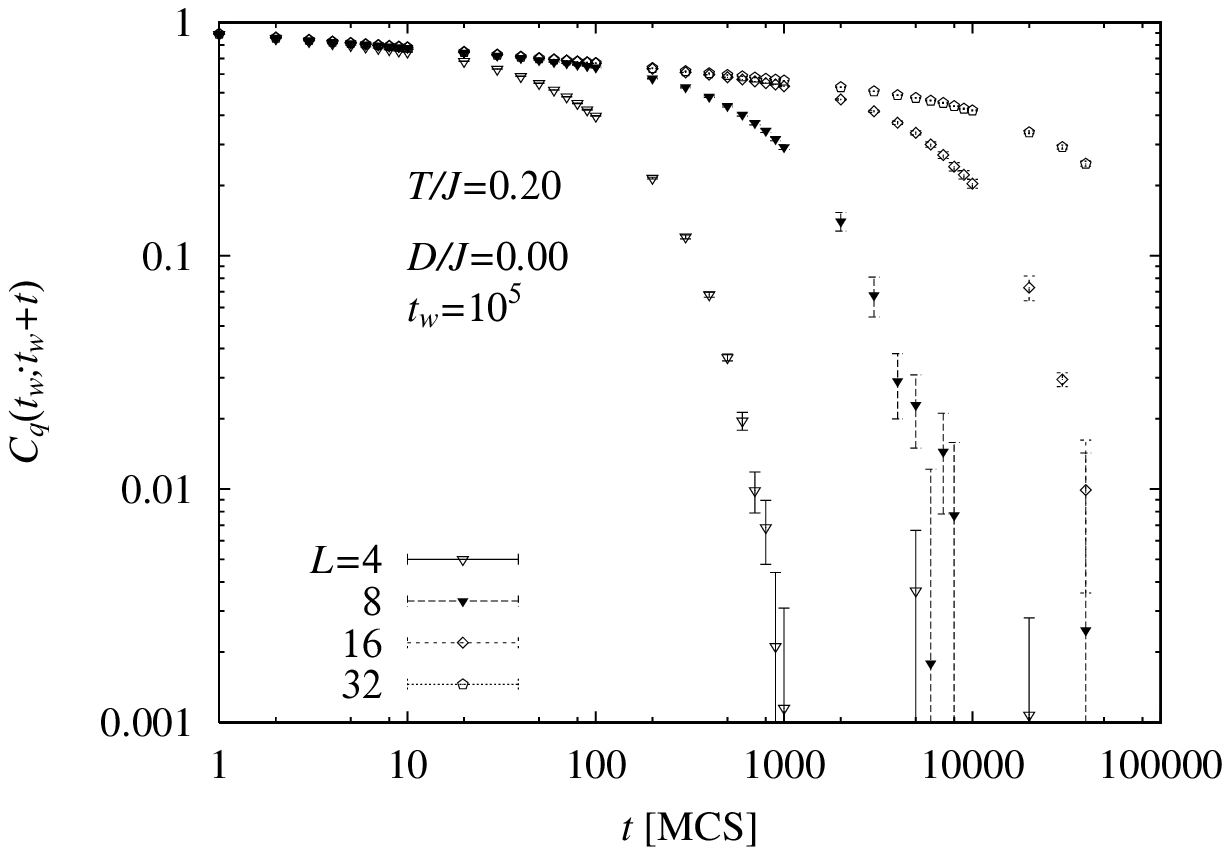}}
\resizebox{\figwidth}{!}{\includegraphics{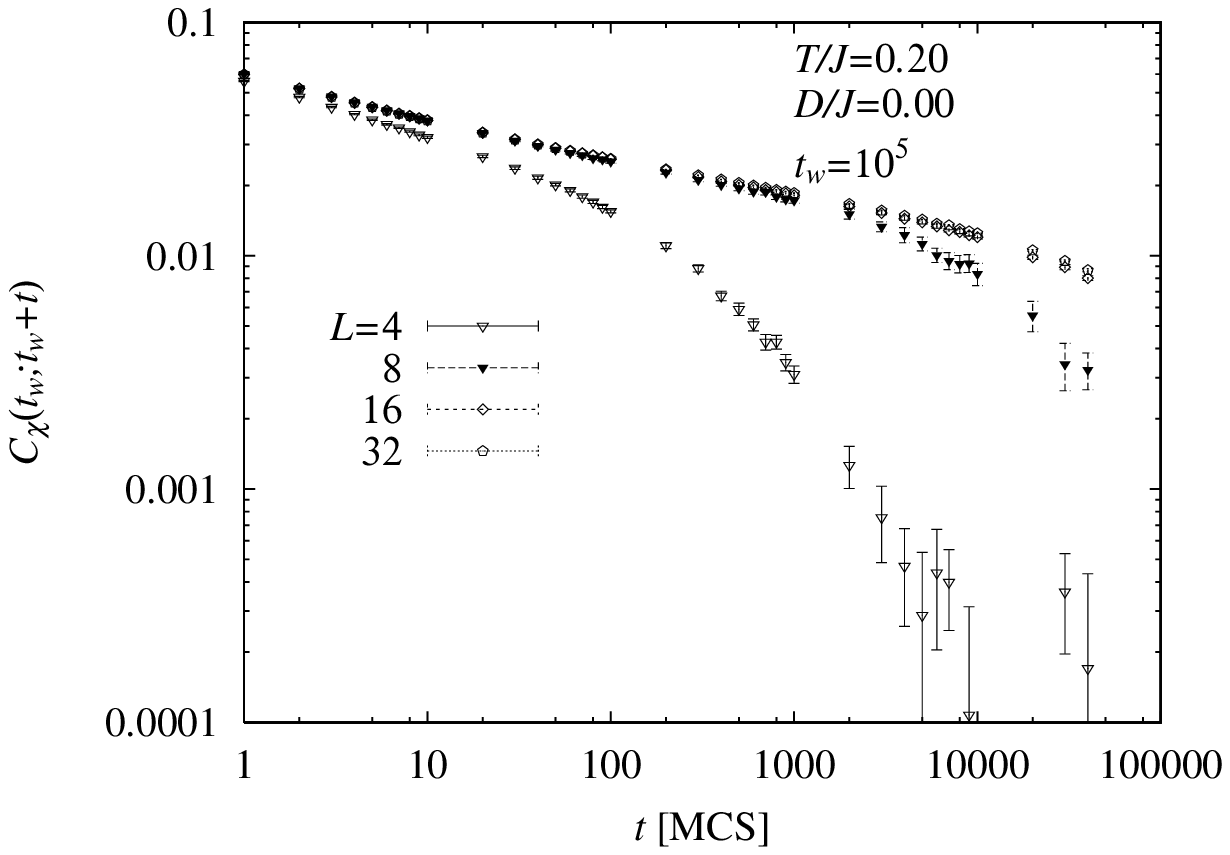}}
\caption{Time dependence of the spin glass (upper) and chiral-glass
 (lower) autocorrelation functions with different sizes and the fixed
 waiting time $t_w=10^5$ in the $3D$
 isotropic Heisenberg SG model at $T/J=0.20$.}
\label{fig3:rawqsize}
\end{figure}

In order to study non-equilibrium dynamics of the model, we
calculate the spin autocorrelation function defined by
\begin{equation}
 C_q(t_w,t)=\frac{1}{N}\sum_i[\langle\vec{S}_i(t_w)\cdot\vec{S}_i(t_w+t)\rangle],
\label{eqn:sec3-spin}
\end{equation}
where $\langle\cdots\rangle$ represents an average over initial
conditions and $[\cdots]$ over the bond disorder. 
In the long time limit ($t\rightarrow\infty$) after
taking the equilibrium limit ($t_w\rightarrow\infty$) the
autocorrelation,  Eq.~(\ref{eqn:sec3-spin}), converges to the  SG 
order parameter originally proposed by Edwards and Anderson\cite{EA75}. 
We also calculate the corresponding chirality autocorrelation
function, which allows us to detect a possible CG transition
from the dynamical point of view. 
It is also defined as an overlap of the local chirality 
\begin{equation}
 C_\chi(t_w,t)=\frac{1}{3N}\sum_i[\langle\vec{\chi}_i(t_w)\cdot\vec{\chi}_i(t_w+t)\rangle],
\label{eqn:sec3-chi}
\end{equation}
where $\chi_{i \mu}$ is the local chirality at the $i$th site and in the
$\mu$th direction  defined by a scalar as 
\begin{equation}
\chi_{i \mu} = \vec{S}_{i+\hat{e}_\mu} \cdot (\vec{S}_i \times \vec{S}_{i-\hat{e}_\mu}) , 
\label{scalar-chirality}
\end{equation}
with $\hat{e}_\mu (\mu =x,y,z)$ being a unit lattice vector along the
$\mu$ axis. 
The CG autocorrelation function in off-equilibrium has not been 
investigated extensively yet, except for Ref.~\onlinecite{Kawamura98}. 

In Fig.~\ref{fig3:rawq}, we show the SG and CG 
autocorrelation functions with the different waiting times $t_w$
against time $t$.  
In non-equilibrium dynamics, temporal correlation functions explicitly depend
on not only $t$ but also $t_w$, which we call the breaking of time
translational invariance.  The longer is the waiting time, the slower 
becomes the relaxation. The curves with shorter waiting times start to
deviate from an equilibrium function at the shorter time scales.  
Such a time scale,  below which the time translational invariance 
holds effectively,  is roughly proportional to the waiting time $t_w$.

In Fig.~\ref{fig3:rawqsize} we present the correlation functions
with different system sizes 
for a fixed waiting time $t_w=10^5$. 
One can clearly see that 
the relaxation curves start to deviate from that of the thermodynamic
limit. 
This implies that finite-size effects yield another characteristic
relaxation time.  
The latter is related to the critical slowing  down which we study in
the present work.

Generally speaking, equilibrium MC simulations for SG models suffer
from extremely slow dynamics and equilibration is hardly realized in
large systems at low temperatures. 
One of the advantages of the non-equilibrium simulation is to
practically avoid  such a very slow equilibration process. 
In the present work,  we focus our attention to the longest relaxation
time of finite-size systems which is expected to diverge with the system
size at and below the transition temperature. 
Our strategy of the non-equilibrium simulation is to take the
equilibrium limit, i.e., $t_w\rightarrow\infty$, after extracting the
characteristic relaxation time under off-equilibrium in finite systems
and then take the thermodynamic limit $L\rightarrow\infty$.

\section{\label{sec:ratio}Dynamical ratio function and relaxation time }
According to the dynamic scaling ansatz\cite{Hohenberg}, 
the autocorrelation function in equilibrium near a second-order phase
transition has the form 
\begin{equation}
 C(t_w=\infty,t_w+t)\sim t^{-\lambda}f\left(t/\tau\right),
\label{eqn:sec3-dscaling}
\end{equation}
where $\tau$ is the relaxation time at a given temperature and $\lambda$
a critical exponent. This form represents  a crossover around
$\tau$ from the short-time critical behavior, $t^{-\lambda}$, to the
long-time relaxation described by $f(t/\tau)$.  
When the temperature approaches to the  critical point $T_c$, the relaxation
time $\tau$ diverges as $\tau(T)\sim |T-T_c|^{-z\nu}$. Thus, 
the autocorrelation function follows a power law  in $t$ at $T_c$.  

One may consider that the critical temperature as well as the critical
exponents associated with the transition can be easily obtained by the
relaxation time near the critical temperature. There are, however, 
several difficulties in obtaining the relaxation time by MC simulations.
One difficulty is due to extremely slow dynamics of SG systems as
mentioned above. 
Another difficulty is that the method of obtaining the relaxation time
requires accurate estimation of a long-time tail of the autocorrelation
function.  One way is to integrate $C(t)$ over time. A contribution
to the integrated relaxation time is dominated by the long-time tail
where $C(t)$ simulated is very small and largely fluctuating. Another
method would be a fitting of $C(t)$ to a relaxation function with the
scaled argument 
$t/\tau$. It is, however, difficult to know the functional form
explicitly, which is often non-exponential as seen in
Fig.~\ref{fig3:rawq}.

\begin{figure}
\resizebox{\figwidth}{!}{\includegraphics{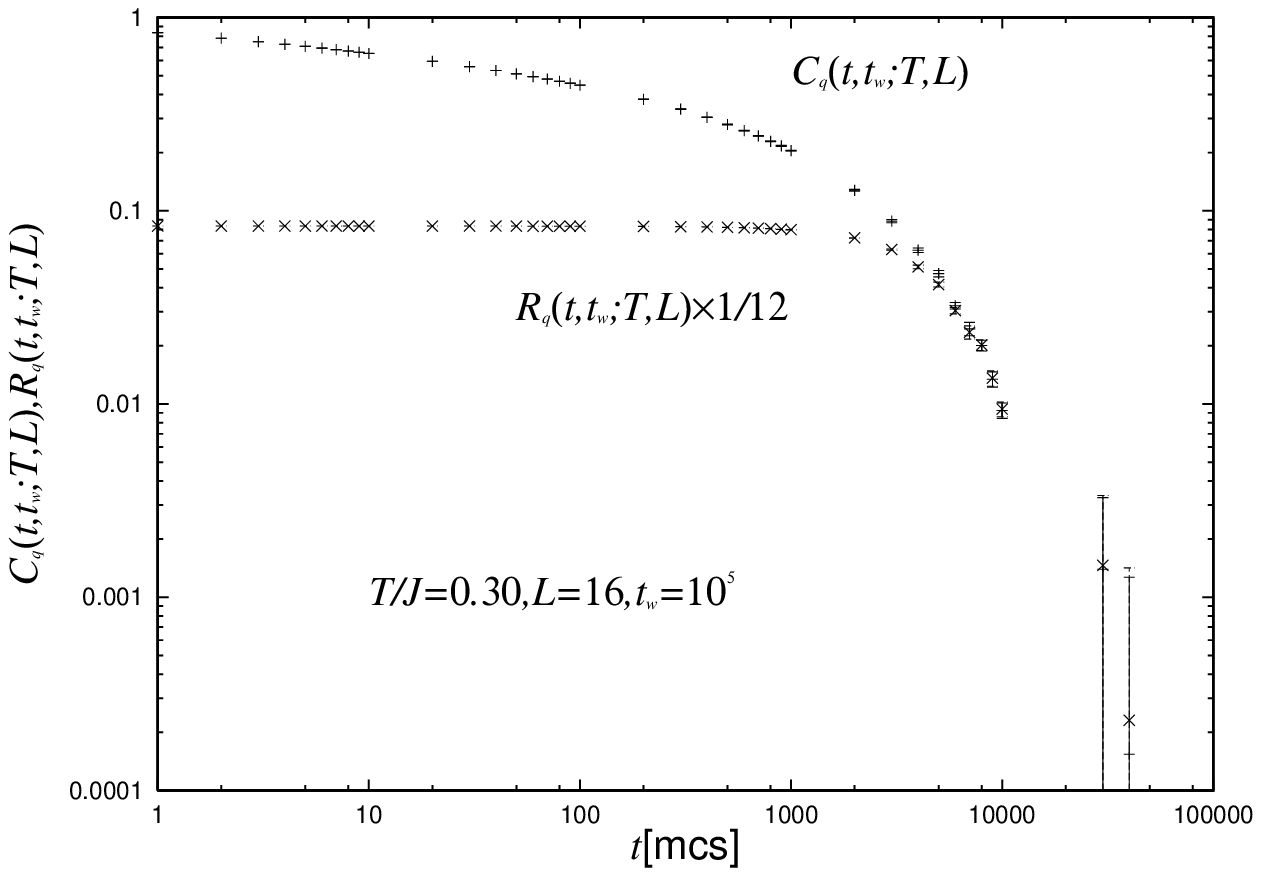}}
\resizebox{\figwidth}{!}{\includegraphics{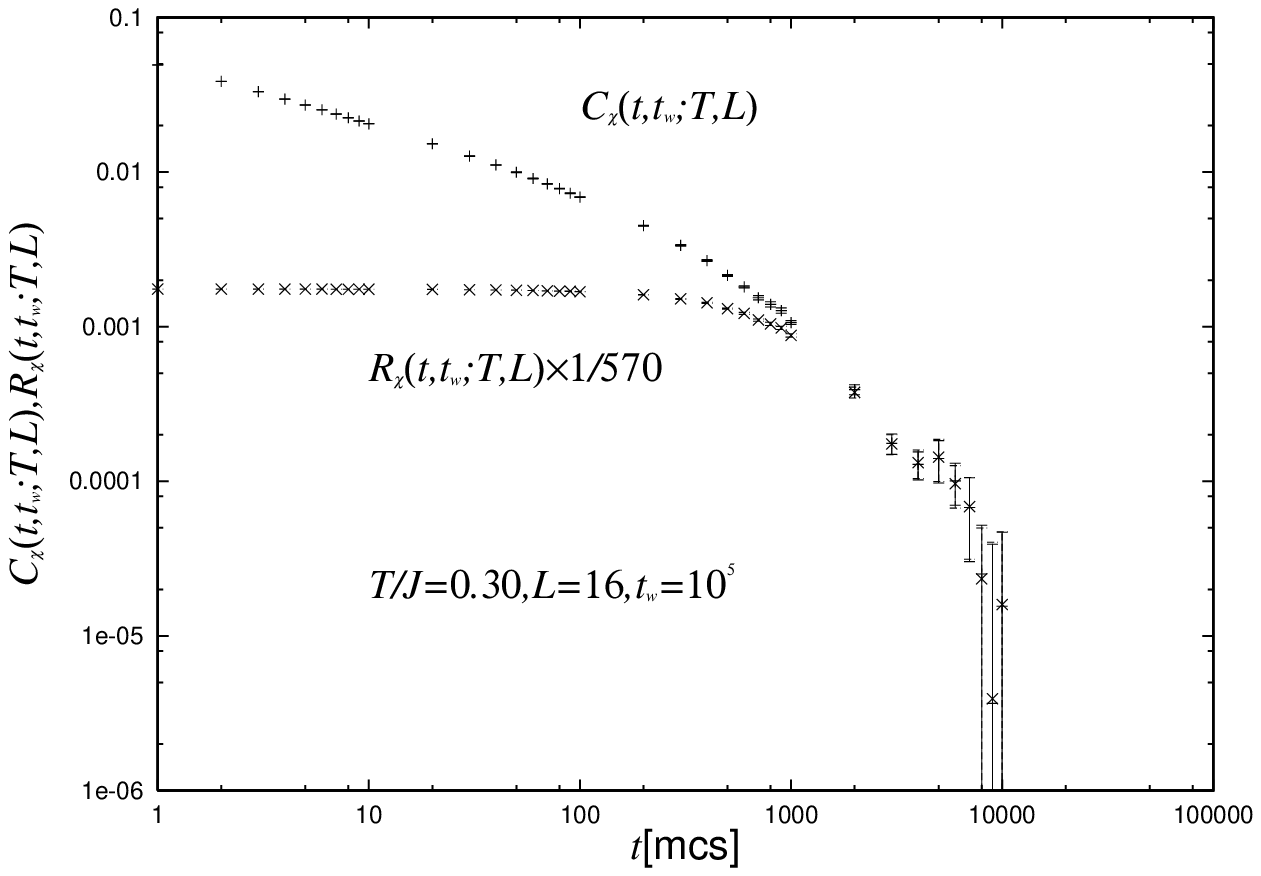}}
 \caption{Double logarithmic plot of the spin-glass  (upper)  and
 chiral-glass (lower)  autocorrelation functions and  ratio functions of
 the isotropic $3D$ Heisenberg spin glass model at $T/J=0.30$. }
\label{fig:fig3-1}
\end{figure}

In order  to avoid the above mentioned difficulties, Bhatt and
Young\cite{Bhatt-Young92} (hereafter referred to as BY) have 
introduced a dimensionless dynamic correlation function 
\begin{equation}
R_q(t_w,t_w+t)= \frac{C_q(t_w,t_w+t)}{\sqrt{ \left[\left\langle \left(
\frac{1}{N} \sum_i \vec{S}_i(t_w) \cdot \vec{S}_i(t_w+t)
\right)^2 \right\rangle \right]}}.
 \label{eqn:ratio-spin}
\end{equation} 
The prefactor in Eq.~(\ref{eqn:sec3-dscaling}) involving a power of $t$
is canceled out by taking the ratio 
of moments. 
A similar dimension-less dynamical function has also been studied in
simple ferromagnets\cite{Jaster,Okabe}. 
The corresponding CG ratio function is defined as 
\begin{equation}
R_\chi(t_w,t_w+t)= \frac{C_\chi(t_w,t_w+t)}{\sqrt{ 
\left[\left\langle \left(
\frac{1}{N} \sum_i^{} \vec{\chi}_i(t_w) \cdot \vec{\chi}_i(t_w+t)
\right)^2 \right\rangle \right]}}.
 \label{eqn:ratio-chirality}
\end{equation} 
Because these ratio functions are dimensionless such as the Binder
parameter in the static case, the dynamical scaling form is expected to be
given by a function of $t/\tau$ as 
\begin{equation}
 R(t)\sim \overline{R}(t/\tau), 
\label{scalR}
\end{equation}
where $\overline{R}(x)$ is a universal scaling function. 
BY\cite{Bhatt-Young92} fixed  the waiting time $t_w$ in
Eq.~(\ref{eqn:ratio-spin}) to $t$ because it was believed 
that equilibration of the autocorrelation function on a given time scale
$t$ needed the waiting time $t_w$ of the same time scale as $t$. 
They studied the dimensionless function of the short range Ising SG and the
mean-field Sherrington-Kirkpatrick models just at the known critical
temperature and successfully obtained
the dynamical critical exponent $z$  
using the finite-size scaling for the relaxation times $\tau(L,T_c)\sim
L^z$.  

In the present study, we systematically investigate the ratio function
both for the SG and CG autocorrelation functions as bivariate
functions of $t$ and $t_w$. Instead of taking the equilibrium limit of
the ratio function or fixing the time scale $t_w\sim t$ studied
previously, we try to take the equilibrium limit $(t_w \rightarrow
\infty )$ of the relaxation time $\tau 
(t,t_w)$ obtained from the ratio function for several waiting times.

\begin{figure}
\resizebox{\figwidth}{!}{\includegraphics{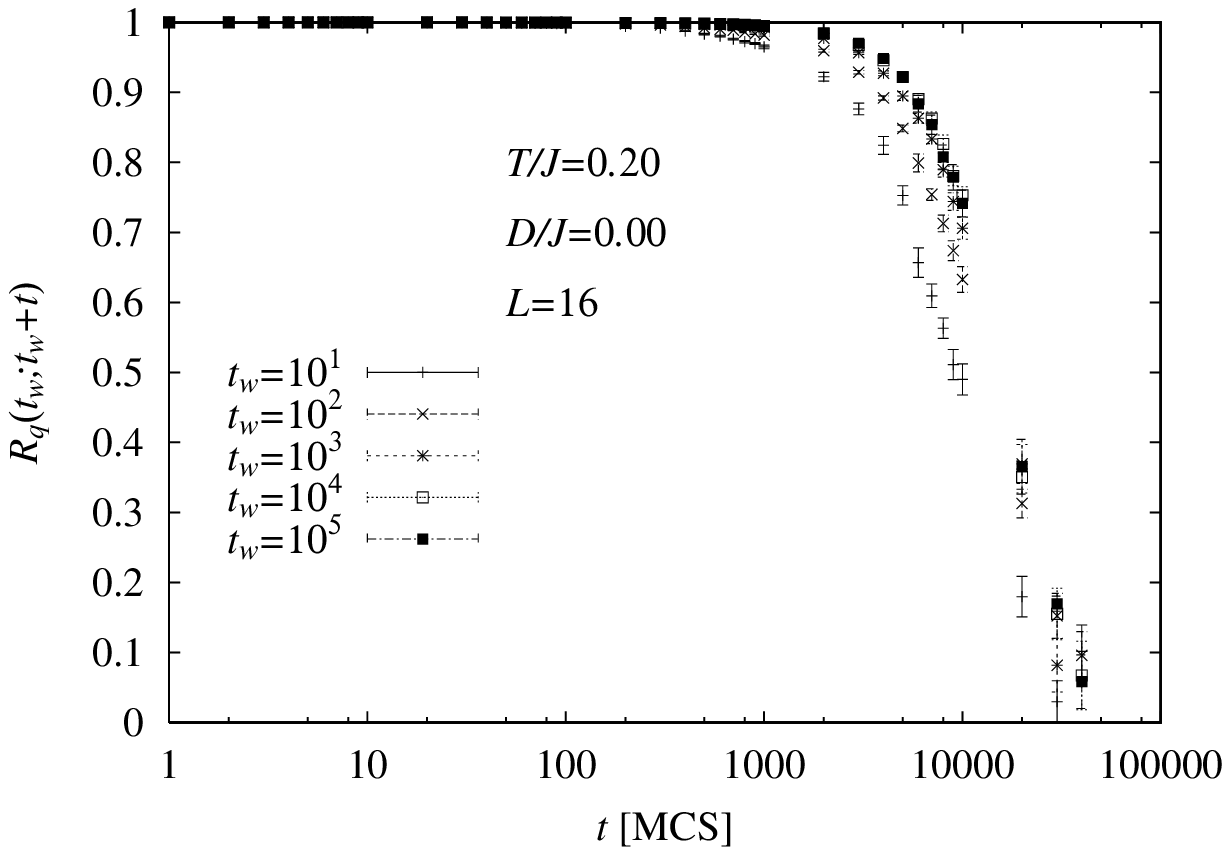}}
\resizebox{\figwidth}{!}{\includegraphics{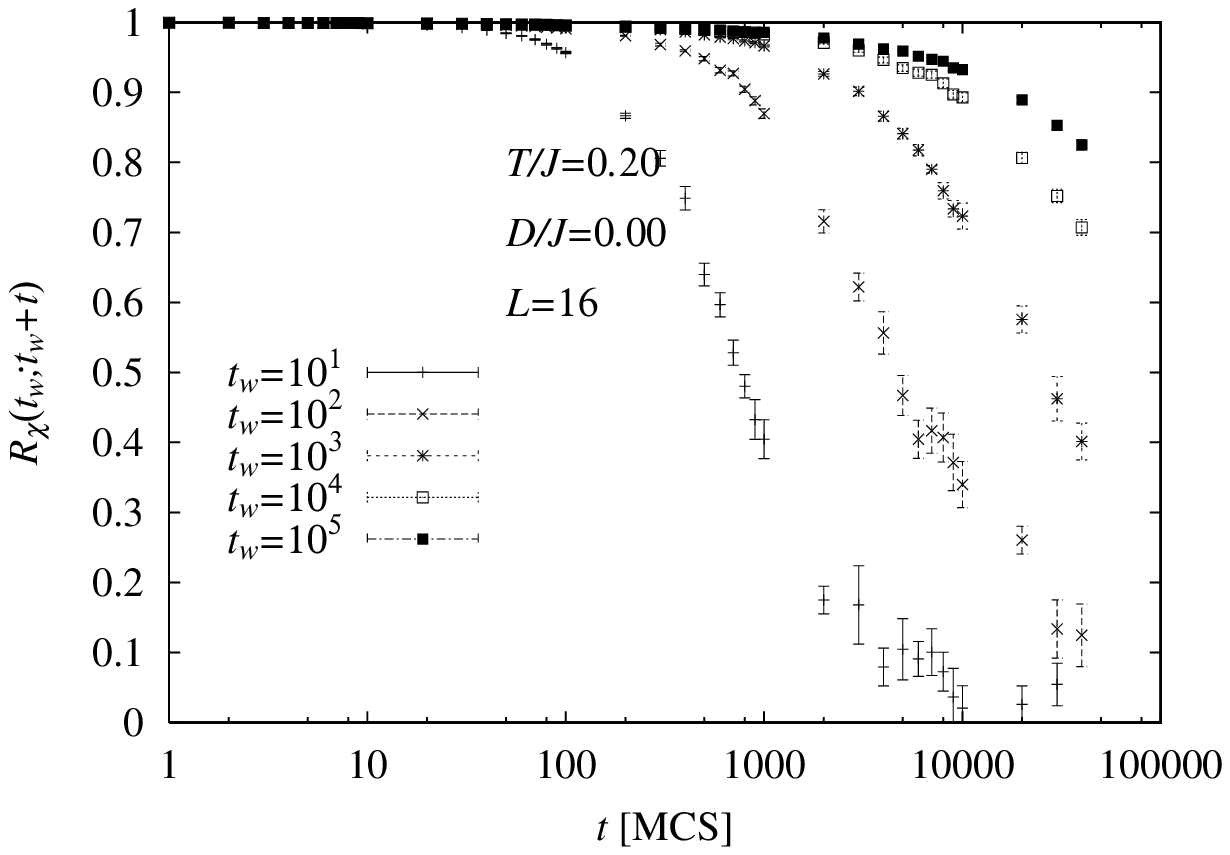}}
\caption{Time dependence of the SG (upper) and CG (lower) ratio
 functions with different waiting times in the $3D$ isotropic Heisenberg
 SG model at $T/J=0.20$.}   
\label{fig3:ratio}
\end{figure}

Equations (\ref{eqn:ratio-spin}) and
(\ref{eqn:ratio-chirality}) are the 
ratio of an odd moment of the autocorrelation function to an even one. 
Consequently, the ratio functions for a given $t_w$ decay from 1 at
$t=0$ to zero as $t$ goes to infinity and they are sensitive to flips or
rotations of the entire system, which are the longest mode of
relaxation. 
Therefore we expect that the ratio functions pick up only slowest
relaxation modes from the whole modes of $C(t)$. 
A typical example of the ratio function, as well as $C(t)$, is presented 
for the SG and CG autocorrelations in Fig.~\ref{fig:fig3-1}. 
As expected, the ratio functions are unity in the
relatively short-time regime, implying that the denominator and
the numerator coincide with each other. 
For the longer-time regime, as shown in the figure, the ratio functions of
spin and chirality 
sector coincide with the tail of the corresponding autocorrelation
function multiplied by  a certain constant, which represents a
statistical weight of
the slowest modes in $C(t)$. 

\begin{figure}
\resizebox{\figwidth}{!}{\includegraphics{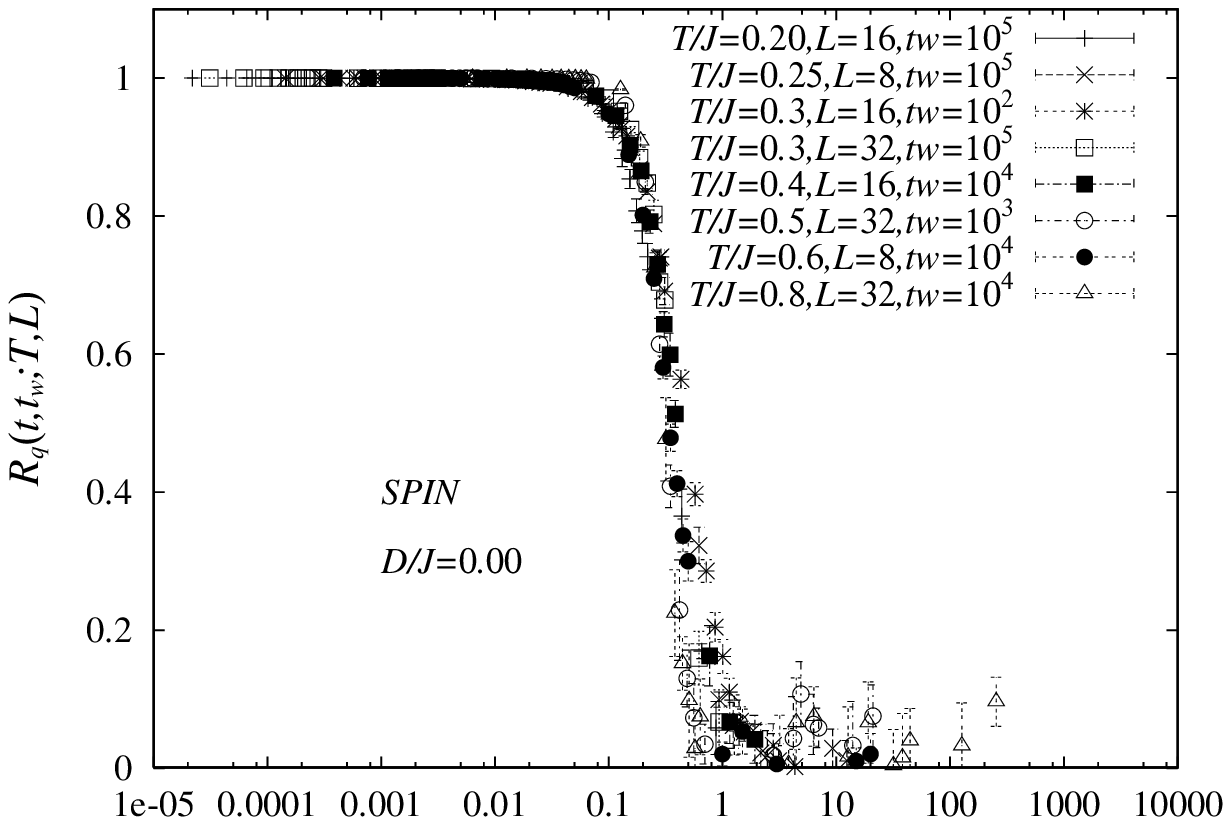}}
\resizebox{\figwidth}{!}{\includegraphics{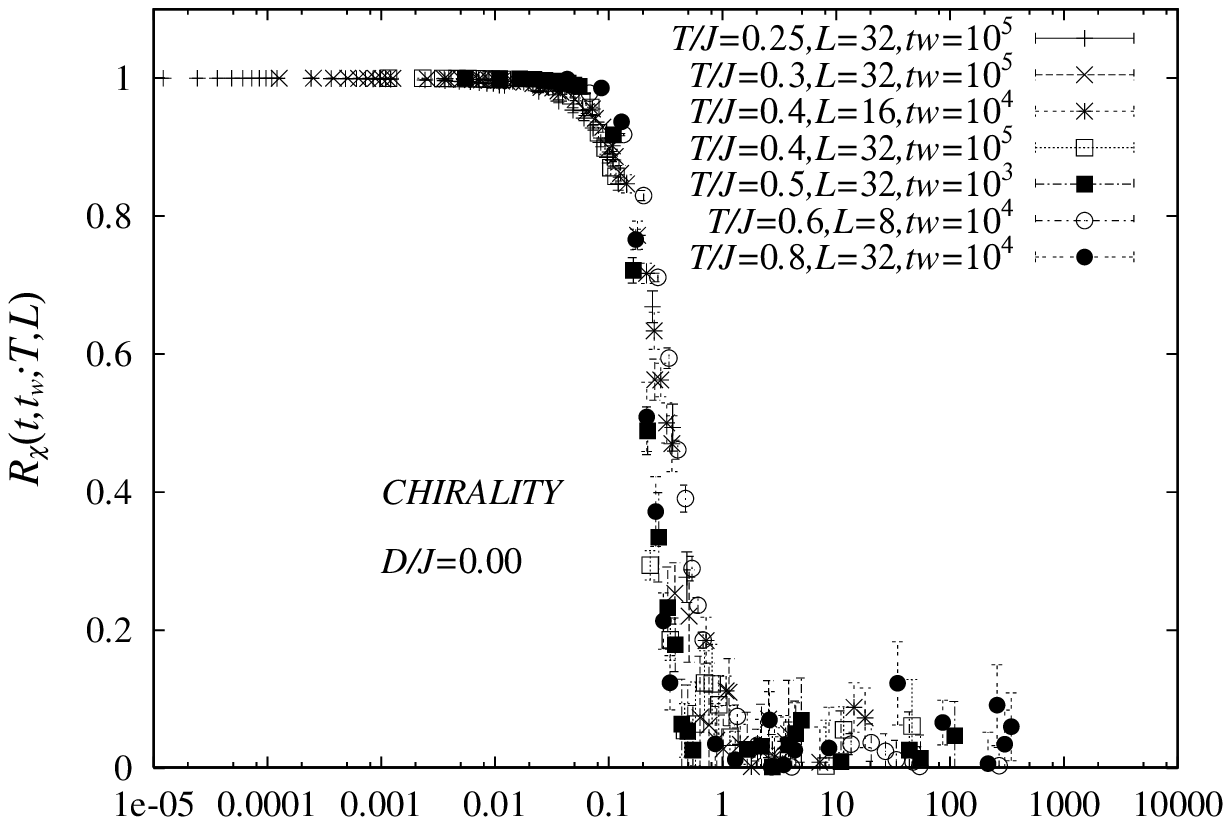}}
 \caption{Scaling plot of the SG(upper) and CG(lower) ratio functions of
 the $3D$ Heisenberg SG model with various temperatures, waiting times
 and system sizes. }
\label{fig:fig3-3}
\end{figure}

When the ratio function $R(t)$ extracts only one
relaxation mode corresponding to the longest mode
in the system, it is expected to show a universal scaling form   
and to be scaled as
\begin{equation}
R(t_w,t_w+t;T,L) \sim R(t/\tau(t_w;T,L)). 
\end{equation}
Here we assume that the ratio functions depend on $t_w, T$ and $L$ only
through the relaxation time. This may be an extension of the dynamical
scaling (\ref{scalR}) to an off-equilibrium situation, which is to be
checked.  
We try to scale the observed ratio functions into a universal scaling
function by appropriately choosing the time constant $\tau$ depending on
$t_w, T$ and $L$. 
Figure~\ref{fig:fig3-3} presents a result of the scaled ratio functions
for $D/J=0$. 
In the figure, one sees that both the SG and CG ratio functions
 with various temperatures, system sizes  and waiting times are almost
 scaled as a  common universal function. 
Note that this finding allows us to compare the absolute values of the
SG and CG relaxation time. 
Using the scaling of the ratio functions in this way, we obtain the
characteristic 
relaxation times $\tau(t_w;T,L)$ for each set of parameters $t_w,T$ and
$L$. 
We then let $t_w$ to go to infinity to obtain the characteristic relaxation
times in {\it equilibrium}. 
We assume that the $t_w$-dependence of the relaxation times is described
as
\begin{equation} 
\tau(t_w;T,L)=\tau (t_w=\infty;T,L) - \alpha t_w ^{-\beta},
\label{tau-tw}
\end{equation}
with $\alpha, \beta$ being some constants and  
$\tau (T,L) \equiv \tau(t_w=\infty;T,L)$ the equilibrium relaxation
time. 

\begin{figure}
\resizebox{\figwidth}{!}{\includegraphics{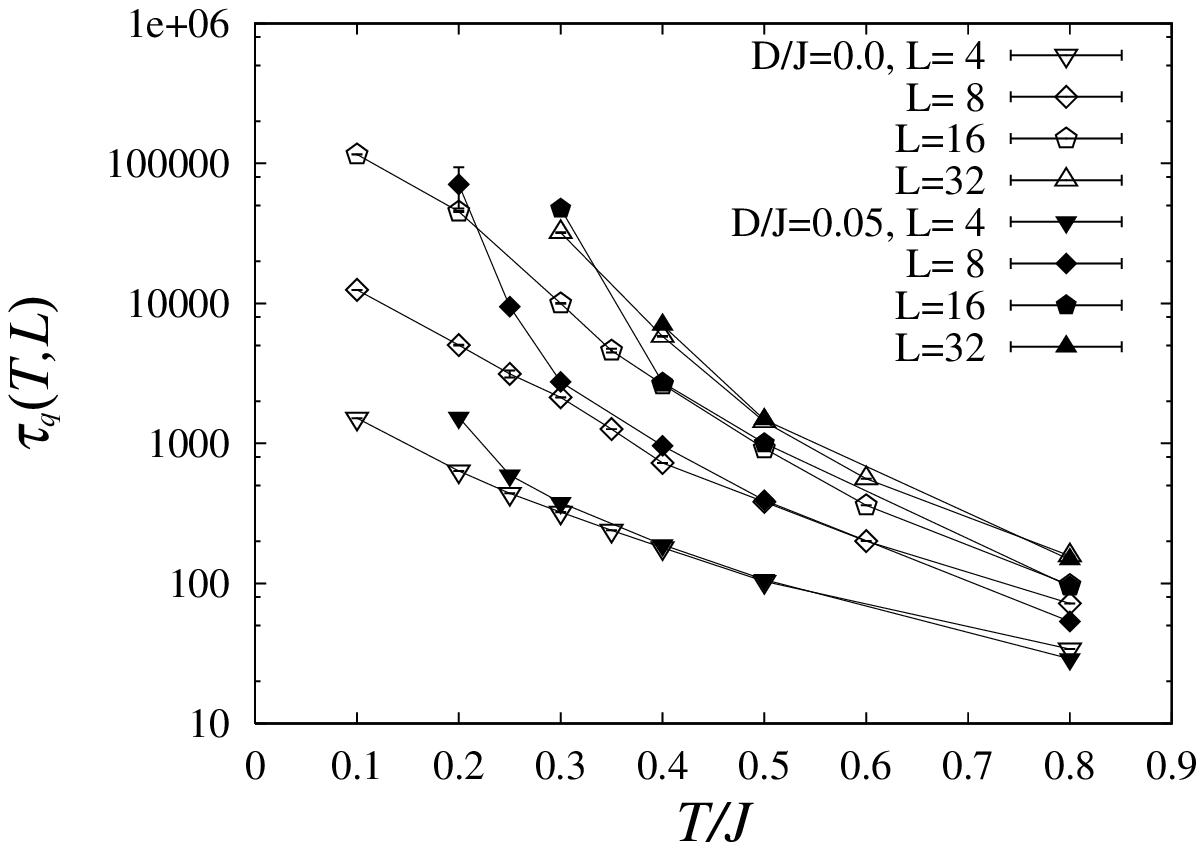}}
\resizebox{\figwidth}{!}{\includegraphics{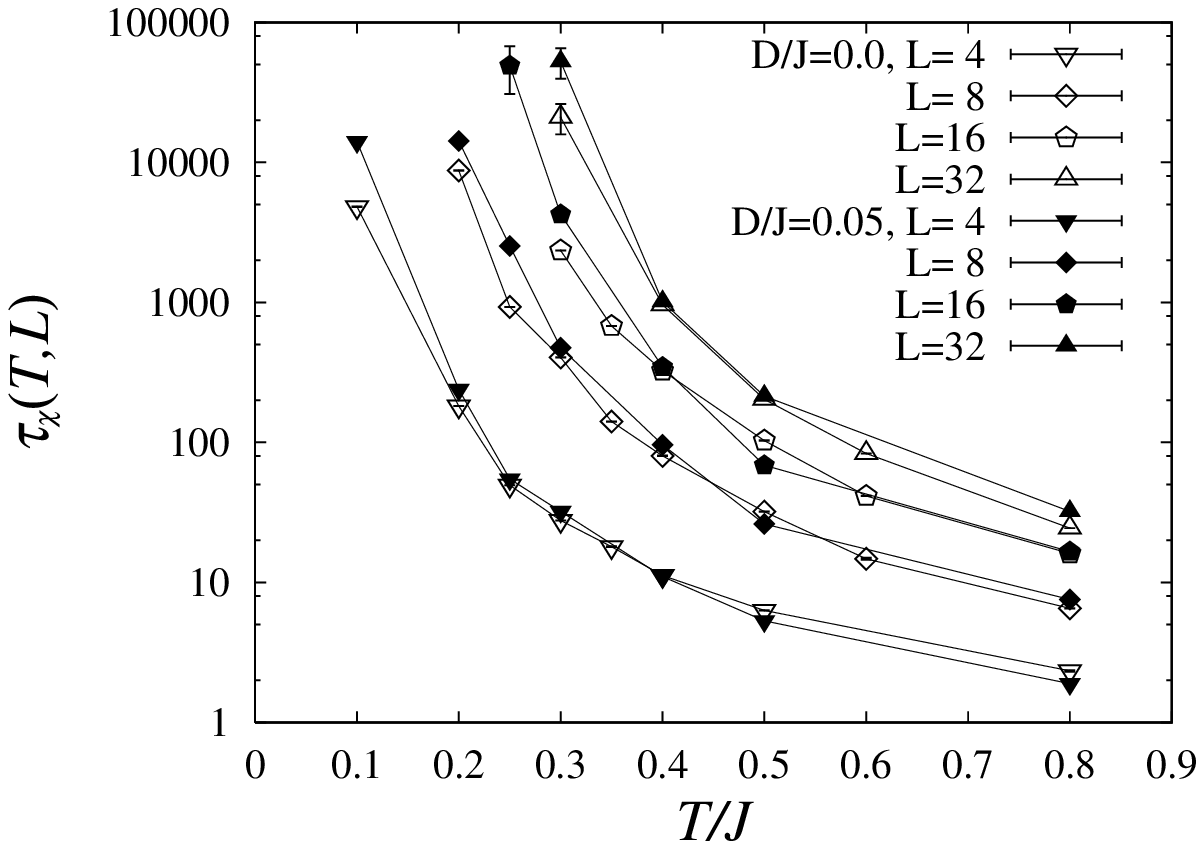}}
 \caption{Temperature dependence of the SG (upper) and CG (lower)
 relaxation times in equilibrium for the isotropic case (open marks) and
 anisotropic case with $D/J=0.05$ (filled marks). 
}
\label{fig:fig3-78}
\end{figure}

Temperature dependences of $\tau(T,L)$ thus extracted are shown in
Fig.~\ref{fig:fig3-78} for the spin (upper) and chiral (lower) degrees of freedom. 
We first note that the CG relaxation time is almost
independent of the strength of the magnetic anisotropy $D$. 
This is natural since the anisotropy breaks the rotational
symmetry, but not the reflection symmetry which is spontaneously broken
by the CG phase transition.   
On the other hand, as seen in Fig.~\ref{fig:fig3-78},
the random anisotropy changes drastically the behavior of the SG
relaxation time. The SG relaxation time of $D/J=0.05$ increases
rapidly around $T/J=0.25$ with decreasing the temperature. 
Such a drastic change of the behavior implies
that the anisotropy mixes the chirality with spin degrees of freedom
as predicted in the chirality scenario.
This is the first observation of the effect of the anisotropy for the SG
relaxation times.

\begin{figure}[]
\resizebox{\figwidth}{!}{\includegraphics{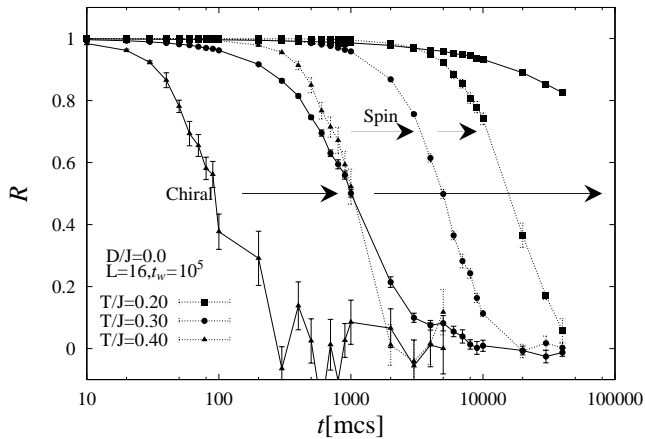}}
 \caption{
Time evolution of the SG(solid) and CG(dotted) ratio functions at $T/J=0.40$,
 $0.30$ and $0.20$ from left to right.  
}
\label{fig:fig3-9}
\end{figure}

\begin{figure}[]
\resizebox{\figwidth}{!}{\includegraphics{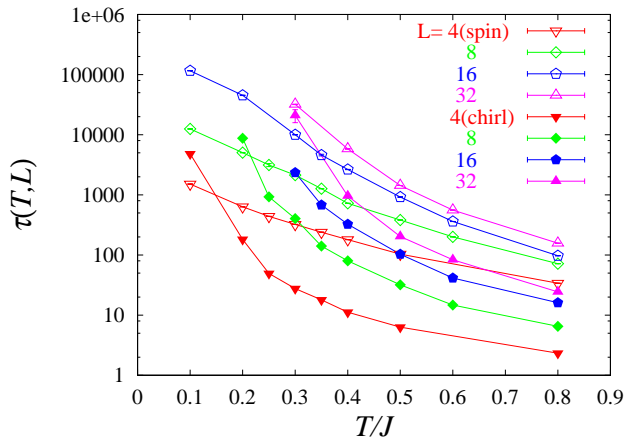}}
 \caption{Temperature dependence of the SG (open) and CG (filled)
 relaxation times in equilibrium  in the isotropic case with different
 sizes. 
}
\label{s-c-tau-equA}
\end{figure}

We can show another evidence that the chirality dominates the phase
transition in the isotropic $3D$ Heisenberg SG model.
In Fig.~\ref{fig:fig3-9}, the time evolution of SG (broken line) and
CG (solid line) ratio functions at $T/J=0.40,0.30$ and $0.20$ with
$L=16$ and $t_w=10^5$ are shown simultaneously. 
The characteristic relaxation time of the CG ratio function  becomes
larger with decreasing temperature much more rapidly than that of the SG
ratio function does and it finally exceeds the SG relaxation time at
$T/J=0.20$.   
This fact strongly suggests that the CG degrees of freedom dominate large-scale
and long-time phenomena at low temperatures, while the SG one dominates
at relatively high temperatures. 
This type of crossover is observed at the first time in the present work. 
We can also see such a crossover in the resulting equilibrium
relaxation time. Figure~\ref{s-c-tau-equA} shows the same data as shown
in Fig.~\ref{fig:fig3-78} but only for the isotropic case $D/J=0$. 
A characteristic temperature at which the two relaxation times cross
each other shifts towards the high-temperature side as the system size
increases. We emphasize that this crossover is not due to an apparent
finite-size effect.   

\section{\label{sec:FSS}Finite-size scaling}
In the present section we discuss the finite-size-scaling analysis of
the equilibrium relaxation time obtained in the previous section.  
According to a standard dynamical finite-size-scaling analysis, the 
scaling form of $\tau(T,L)$ is given by
\begin{equation}
 \tau(T,L)\sim L^z \overline{\tau}\left((T-T_c)L^{1/\nu}\right),
\label{eqn:fss}
\end{equation}
where $z$ is the dynamical critical exponent and $\nu$ the
correlation-length exponent.  
The scaling function becomes a constant just at the critical
temperature, leading the finite-size scaling form, $\tau(T_c,L)\sim
L^z$, previously studied\cite{Bhatt-Young92}. 
We discuss asymptotic behavior of the scaling function
$\overline{\tau}(x)$. 
In a large-system-size limit above $T_c$, i.e., $x\gg 1$, the
scaling of Eq.~(\ref{eqn:fss}) should recover an ordinary scaling form, 
\begin{equation}
 \tau(T,L=\infty)\sim (T-T_c)^{-z\nu}.  
\label{eqn:fsstau}
\end{equation}
This leads that the scaling function behaves as 
\begin{equation}
 \overline{\tau}(x)\sim x^{-z\nu}
\label{eqn:asymptotic1}
\end{equation}
at $x\gg 1$.  
Assuming that the transition takes place at a finite temperature,  
the opposite limit, i.e. $-x\gg 1$, can be also argued in the isotropic
case where the system would entirely behave like a rigid magnet. 
For the SG ordering in $D/J=0$, such a  system
rotates globally as coherent diffusive motion with the 
relaxation time $\tau_D$ proportional to the bulk volume. 
This implies that the asymptotic scaling form of the SG relaxation
time in $D/J=0$ is 
\begin{equation}
 \overline{\tau}(x)\sim |x|^{(d-z)\nu}, 
\label{eqn:asymptotic2}
\end{equation}
in the limit $-x\gg 1$  with $d$ being the dimensionality. 
The asymptotic scaling function for the CG relaxation time, however,  
would follow an exponential activated type because of the discrete
nature of chirality. 

The situation is complicated in a system with $T_c=0$.  
In one case where the conventional form (\ref{eqn:fss}) holds with
$T_c=0$  as in the standard second-order phase transition,
the scaling function in the asymptotic limit at  $x\ll1$,
$\overline{\tau}(x=0)$,  becomes a constant. On the other hand,  
Yoshino and Takayama\cite{Yoshino93} proposed a modified scaling form
which was a 
combination of the standard form (\ref{eqn:fss}) with $T_c=0$ and the
diffusion relaxation time, $\tau_D\sim L^d/T$. This form is compatible
with (\ref{eqn:fss}) and (\ref{eqn:asymptotic2}) in the limit $x\ll1$ if
$(d-z)\nu=-1$ holds. In this case, the scaling exponents, $z$ and $\nu$,
are no longer independent parameters. 

\subsection{Isotropic case: $D/J=0$ }
\begin{figure}
\resizebox{\figwidth}{!}{\includegraphics{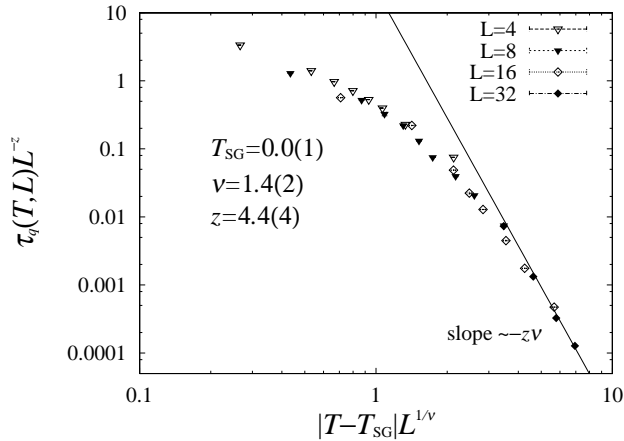}}
\caption{Finite-size scaling plot of the SG relaxation time in
 the isotropic $3D$ Heisenberg SG model. The parameters of the scaling
 are as follows : 
$T_{\rm SG}=0.0(1)$, $\nu=1.4(2)$ and $z=4.4(4)$.
The solid line represents the expected asymptotic behavior for large
 scaling regime.  
}
\label{fig:scale-sgtau-d00}
\end{figure}

The result of the finite-size scaling analysis mentioned
above is demonstrated for the SG relaxation times of the isotropic
Heisenberg SG model in Fig.~\ref{fig:scale-sgtau-d00}. 
As can be seen from the figure, the data points are found to collapse
into the universal scaling function. 
From the scaling we estimate
$T_{\rm SG}=0.0(1)$, $\nu=1.4(2)$ and $z=4.4(4)$. 
Our result for the SG transition temperature is roughly consistent with
the zero-temperature phase
transition\cite{Banavar,McMillan85,Olive86,Yoshino93,Matsubara91},
although we could not completely 
exclude the possibility of the SG phase transition at a very low but
finite temperature.
Assuming the zero-temperature transition, the
scaling results suggest that relaxation times diverge with a power law
of $T$ as $T$ goes to 0, but not an exponential divergence which occurs at 
the lower critical dimensions such as the two-dimensional Ising SG model. 

We expect that the scaling function converges to a finite value in the
limit $x\rightarrow 0$ as in the standard second-order phase transition,
though we don't have enough data to confirm this convergence. 
The scaling plot may be consistent with a modified scaling from $x\gg 1$
to $x\ll 1$ proposed by Yoshino and Takayama\cite{Yoshino93}, but then 
temperature dependence of the 
relaxation time at low temperatures, $\tau\sim T^{(d-z)\nu}$ with $z$
and $\nu$ obtained above does not coincide with that expected from a
simple diffusive relaxation time proportional to $1/T$. 
A further investigation is needed to clarify this point. 

The present estimate for $\nu$ is compatible with the previous ones in
other works;   
$\nu=1.54(19)$\cite{McMillan85} and $2.0(2)$\cite{Kawamura92} by a
domain-wall renormalization group calculation at $T=0$,  
$\nu=1.35(5)$ by a MC simulation\cite{Matsubara91}, 
$\nu\sim 1.6$ by an equilibrium dynamics\cite{Yoshino93} and 
$\nu\sim 1.14$ by a MC simulation\cite{Olive86}. 

\begin{figure}
\resizebox{\figwidth}{!}{\includegraphics{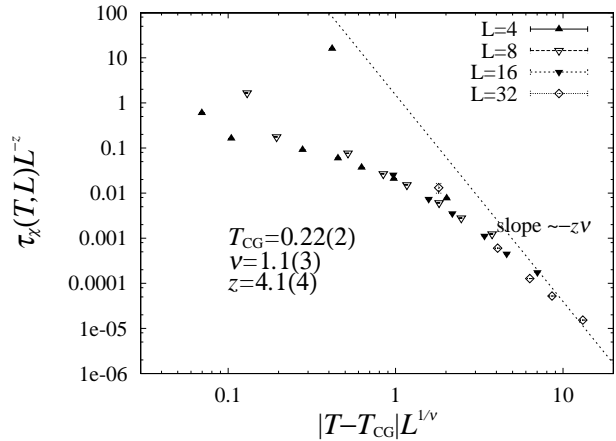}}
\caption{Finite-size scaling plot of the CG relaxation time in
 the isotropic $3D$ Heisenberg SG model. The parameters of the scaling
 are as follows;  
$T_{\rm CG}=0.22(2)$, $\nu=1.1(3)$ and $z=4.1(4)$.
}
\label{fig:scale-cgtau-d00}
\end{figure}

\begin{figure}
\resizebox{\figwidth}{!}{\includegraphics{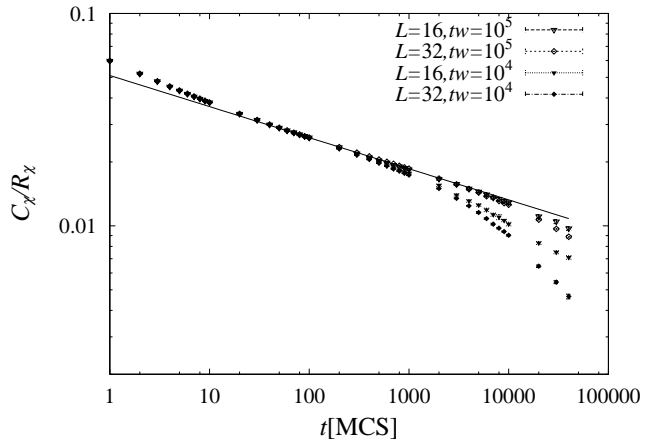}}
 \caption{
Time dependence of $C_\chi/R_\chi$ for the CG at $T/J=0.20$. 
}
\label{fig:cr}
\end{figure}

We have also analyzed the CG relaxation time by a similar
scaling. 
The best scaling shown in Fig.~\ref{fig:scale-cgtau-d00} is obtained by
the following estimates: 
$T_{\rm CG}=0.22(2)$, $\nu=1.1(3)$ and $z=4.1(4)$. 
The estimate for $T_{\rm CG}$ is consistent with the recent equilibrium  MC
calculations\cite{Imagawa01,Hukushima02}. 
The present estimate 
for $\nu$ is rather close to the previous MC estimate for the Gaussian
distribution, $\nu\sim 1.2$\cite{HukushimaKawamura00}.

At the critical temperature, the autocorrelation
function $C_\chi$ and also $C_\chi/R_\chi$ follow the algebraic decay
(\ref{eqn:sec3-dscaling}) with the exponent $\lambda=\beta/z\nu$. 
In Fig.~\ref{fig:cr} we show $C_\chi/R_\chi$ as a function of time at
$T/J=0.20$ close to the CG transition temperature. 
The data still depend on both the system size and the waiting time. As a
preliminary study, we estimate the exponent $\lambda=0.23(5)$ from the
slope of the plot, where the numerical error comes from
the uncertainty of the critical temperature. 
This value combined with $z$ and $\nu$ by the finite-size scaling leads
to another exponent, $\beta\sim 1.0$, which is consistent with the
previous work\cite{Kawamura98}. 

As mentioned in the introduction, the dynamical phase transition which
is separated from the static one can exhibit at a finite temperature in
some one-step RSB systems. 
It depends on whether the order parameter continuously appears or
discontinuously jumps up at the static transition as the temperature
decreases.   
Examples of the discontinuous one-step RSB transition are the
mean-field $p-$spin glass with $p>2$ and the mean-field $q$-state Potts
glass with $q>4$, while those of the continuous one-step RSB transition
are the mean-field $q$-state Potts glass with $2.8<q\leq 4$. 
Our estimation of the CG transition temperature by the dynamical
finite-size scaling is consistent with the static transition
temperature. This fact means that the CG phase belongs to the latter
class of the one-step RSB.

\subsection{Anisotropic case : $D/J=0.05$  }

\begin{figure}
\resizebox{\figwidth}{!}{\includegraphics{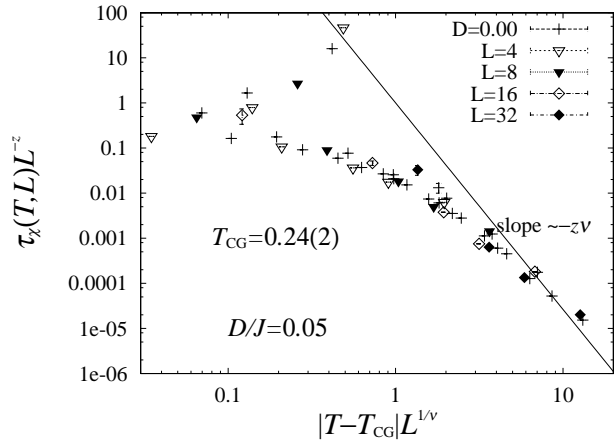}}
\caption{Finite-size scaling plot of the CG relaxation time in
 the $3D$ Heisenberg SG model with $D/J=0.05$. The parameter of the
 scaling  is the CG transition temperature $T_{\rm CG}=0.24(2)$. 
The scaling data of  $D/J=0$ shown in Fig.~\ref{fig:scale-cgtau-d00}
 are also plotted with the cross mark. 
}
\label{fig:scale-cgtau-d05}
\end{figure}

The finite-size scaling plots of the CG relaxation times in 
the anisotropic $3D$ Heisenberg SG model is shown in
Fig.~\ref{fig:scale-cgtau-d05}. 
In the analysis we regard the CG critical temperature as 
a unique adjustable parameter, and obtain $T_{CG}=0.24(2)$. 
As seen in Fig.~\ref{fig:scale-cgtau-d05}, the scaling analysis works
well for both $D/J=0.0$ and $0.05$ with the common $z$ and $\nu$. 
This implies that the critical exponents associated with the CG critical
phenomena as well as the scaling function are independent of the
anisotropic term.  
This finding of our analysis is consistent with the expectation that 
the anisotropic interactions are expected not to affect the CG symmetry.  

We believe that the SG phase transition also occurs in the anisotropic
case  at the same temperature as the CG transition
temperature.  In fact, we have found in Fig.~\ref{fig:fig3-78} that the
anisotropic term significantly enhances the SG relaxation time at low
temperatures. Unfortunately, however, such enhanced relaxation time
appears only in larger sizes and at lower temperatures. Therefore, these
data points are not enough to test the finite-size scaling. 

\section{\label{sec:summary}Discussions}
We have studied dynamical critical phenomena in the $3D$
Heisenberg SG 
models using Monte Carlo simulations. 
Our results support the chirality scenario for
realistic SG transitions. 
It is confirmed that in the fully isotropic model at low temperatures,
the CG relaxation time exceeds the SG relaxation time. 
This result strongly suggests that long-time or large scale behavior is
 dominated by the CG ordering. 
The finite-size-scaling analysis  of the relaxation time
supports that the CG phase transition with the diverging relaxation time
occurs  at a finite temperature, while the SG phase transition takes
place at a lower temperature, consistent with even at zero temperature
as previously studied\cite{Banavar,McMillan85,Olive86,Yoshino93,Matsubara91}. 
In other words, the spontaneous breaking of discrete $Z_2$
symmetry  likely occurs  with preserving the proper rotational symmetry
$SO(3)$.  
The SG and CG transition temperatures  of our result are
significantly separated with each other within numerical accuracy. 
This is the first support for the basic assumption of the chirality
scenario from a dynamical point of view.

We have also studied an anisotropic $3D$ Heisenberg SG model in the
similar way. 
It is found that by introducing the weak random anisotropy, the SG
relaxation times in the large system sizes increase at low temperatures,
similar to that observed in the CG relaxation time, although 
a finite-temperature SG phase transition cannot be directly determined
by the finite-size-scaling analysis. 
Our finite-size-scaling analysis of the CG relaxation time strongly
suggests, on the other hand, that the CG phase transition with the
anisotropic interactions belongs to the same universality class as the
isotropic case. When the SG phase transition in the anisotropic case is
intrinsically dominated by the SG order parameter but not the CG order
parameter, the observed CG critical behavior should be affected by the
SG transition and it should be different from the fully isotropic case. 
Therefore, from this point of view,  our result is consistent with the
chirality mechanism where the phase transition is dominated by the CG
order parameter.  
In order to further clarify the nature of phase transition in the
anisotropic case, however, directly measurements of the SG phase
transition and its critical exponents are required.

 \begin{figure}
 \resizebox{\figwidth}{!}{\includegraphics{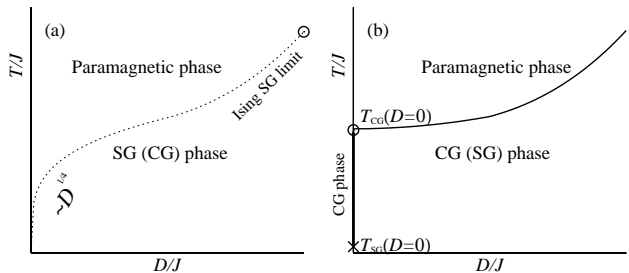}}
 \caption{A schematic picture of a phase diagram in the standard
  scenario (left) and the chirality  scenario (right).}
 \label{phase}
 \end{figure}

A standard scenario\cite{Morris}  based on the zero-temperature SG
transition in the isotropic point says that the SG transition
continuously approaches to zero in the zero anisotropy limit as $T_{\rm
SG}(D)\propto D^{1/4}$ ; see Fig.~\ref{phase}(a). 
Because the anisotropy reduces the symmetry of Hamiltonian, 
the universality class changes from the isotropic SG to the anisotropic
SG. The latter is expected to be in the class of the Ising SG. 
On the other hand, according to the chirality
mechanism\cite{Kawamura92}, the random magnetic anisotropy $D$ 
plays a role of mixing the CG long-range order with the SG one. 
The SG transition temperature in the ideal zero anisotropy limit is
predicted to coincide with $T_{\rm CG}(D=0)$ but not $T_{\rm SG}(D=0)$,
as shown in Fig.~\ref{phase}(b). 
This may be true wherever $T_{\rm CG}>T_{\rm SG}$ at the isotropic limit
even with $T_{\rm SG}(D=0)>0$. 
Once the CG has a long-range order, even the small anisotropy $D$
coherently breaks the rotational symmetry of the system. 
Consequently, the SG critical behavior observed in the anisotropic model
is governed by the fixed point of the CG transition of the isotropic
model. 
The two scenarios can in principle be distinguished by the phase
boundary near the isotropic limit and the universality class along the
phase boundary. 
Our findings that the CG universality class is not affected by the
anisotropy $D$ and that $T_{\rm SG}<T_{\rm CG}$ in the isotropic system
prefer the chirality scenario.

In summary, 
spin-glass and chiral-glass orderings in three-dimensional Heisenberg
spin glass model are studied from the point of view of dynamics by
off-equilibrium Monte Carlo simulations. 
Our results support the chirality scenario for the SG phase transition
in three dimensions.

\section*{Acknowledgements}
The authors would like to thank H.~Kawamura for useful discussions. 
This work is supported by 
a Grant-in-Aid for Scientific Research Program (\# 12640367) 
and that for the Encouragement of Young Scientists(\# 13740233) 
from the Ministry of Education, Science, Sports, Culture and
Technology of Japan. 
The numerical calculations were mainly made by use of the Hitachi
SR8000/60  at the Supercomputer Center, Institute for Solid State
Physics, the University of Tokyo.

\end{document}